# Growth of C-Axis Textured AlN Films on Vertical Sidewalls of Silicon Micro-Fins

Mehrdad Ramezani, *Student Member, IEEE*, Valeriy Felmetsger, Nicholas Rudawski, and Roozbeh Tabrizian, *Member, IEEE*

*Abstract*— A fabrication process is developed to grow c-axis textured aluminum nitride (AlN) films on the sidewall of single crystal silicon (Si) micro-fins to realize fin bulk acoustic wave resonators (FinBAR). FinBARs enable ultra-dense integration of high quality-factor ($Q$) resonators and low-loss filters on a small chip footprint and provide extreme lithographical frequency scalability over ultra- and super-high-frequency regimes. Si micro-fins with large aspect ratio are patterned and their sidewall surfaces are atomically smoothened. The reactive magnetron sputtering AlN deposition is engineered to optimize the hexagonal crystallinity of the sidewall AlN film with c-axis perpendicular to the sidewall of Si micro-fin. The effect of bottom metal electrode and surface roughness on the texture and crystallinity of the sidewall AlN film is explored. The atomic-layer-deposited platinum film with (111) crystallinity is identified as a suitable bottom electrode for deposition of c-axis textured AlN on the sidewall with c-axis orientation of 88.5°±1.5° and arc-angle of ~12° around (002) diffraction spot over film thickness. 4.2 GHz FinBAR prototype is implemented showing a $Q$ of 1,574 and effective electromechanical coupling ($k_{eff}^2$) of 2.75%, when operating in 3rd width-extensional resonance mode. The lower measured $Q$ and $k_{eff}^2$ compared to simulations highlights the effect of granular texture of sidewall AlN film on limiting the performance of FinBARs. The developed c-axis textured sidewall AlN film technology paves the way for realization and monolithic integration of multi-frequency and multi-band FinBAR spectral processors for the emerging carrier aggregated wireless communication systems.

## I. INTRODUCTION

DENSE integration of multi-frequency and multi-band acoustic spectral processors is essential for realization of the emerging ultra-wideband mobile communication systems that operate based on dynamic configuration and carrier aggregation. These systems require a large array of filters with frequencies extended over ultra- and super-high-frequency regimes to enable frequency hopping and spread-spectrum data communication [1-3]. Realization of a single-chip multi-frequency filter solution requires the development of an acoustic resonator technology with true lithographical frequency scalability. A novel acoustic resonator architecture that can provide such requirement is the recently introduced Fin Bulk Acoustic Resonator (FinBAR) [4-8]. The FinBAR is created from the integration of piezoelectric transducer stack on the sidewall of a single crystal silicon (Si) micro-fin and benefits from the large longitudinal piezoelectric effect (i.e. $d_{33}$) to excite the width-extensional bulk acoustic resonance modes. The major advantage of the FinBAR is its frequency tailorability by the lithographically defined Si fin width. By changing the Si fin width, the FinBAR frequency can be tailored over a wide spectrum [5]. This potentially enables single-chip integration of a large array of filters with frequencies extended over the ultra- and super-high-frequency regimes. In practice, however, the realization of high-performance FinBARs is tied to the creation of sidewall piezoelectric transducers with proper texture and crystallinity. This is particularly challenging, considering the limitations of the conventional piezoelectric film deposition techniques for three-dimensional integration of the transducer. The available techniques are developed for growth of the planar films on top of smooth and lattice-matched substrates, and are not applicable for crystalline growth of textured films on the sidewalls of patterned microstructures.

In this paper a novel process is presented based on the reactive magnetron sputtering of densely textured AlN films with perpendicular c-axis on the sidewall of Si micro-fins. Various experiments are performed to identify the challenges in the integration of sidewall AlN transducers. The resulting films are evaluated using scanning and transmission electron microscopy (SEM and TEM) characterization, and FinBAR prototypes. An optimized process is developed and presented for the growth of densely textured AlN piezoelectric transducers with perpendicular c-axis on the sidewall of Si micro-fins.

## II. EXPERIMENTAL PLAN

To facilitate investigation of the process for growth of c-axis textured sidewall AlN films, micro-fin structures are created on (110) Si substrate. This enables creation of Si micro-fins with (111)-oriented sidewall surface, which is the preferred orientation for the growth of hexagonal AlN and also enables realization of high quality factor ($Q$) FinBARs with reduced support loss [6]. Si micro-fins with various aspect ratio and distribution density are fabricated using Bosch DRIE process. Due to the essence of this process, which consists of a finite number of isotropic etch and passivation cycles, the sidewall surface of micro-fins suffers from large roughness and scalloping. Figure 1(a) shows the cross-sectional SEM of the Si micro-fins after DRIE, highlighting a scallop depth of 23 nm. Figure 1(b) shows the sidewall topography measured using a Bruker Optical Profilometer tool, highlighting a surface roughness of ~28 nm in root-mean-square (rms). Achieving a high quality AlN films requires sub-1 nm surface roughness [9][10]. To reduce the 28nm sidewall roughness after the DRIE process, a hydrogen ($H_2$) annealing step is performed at 1100°C is used to smoothen sidewalls [11]. Figure 1(c-e) shows the micro-fins after $H_2$ annealing process. A set of seven distinct experiments are performed on the substrates with smoothened micro-fins, to identify the effect of processing parameters and transducer stack details on the quality of sidewall AlN films. While the details of these experiments are elaborated in the supplementary document, this paper presents the results of four representative experiments (wafers 1-4) that highlight the challenges in achieving textured sidewall AlN films, and the optimized process that yields the desirable quality.

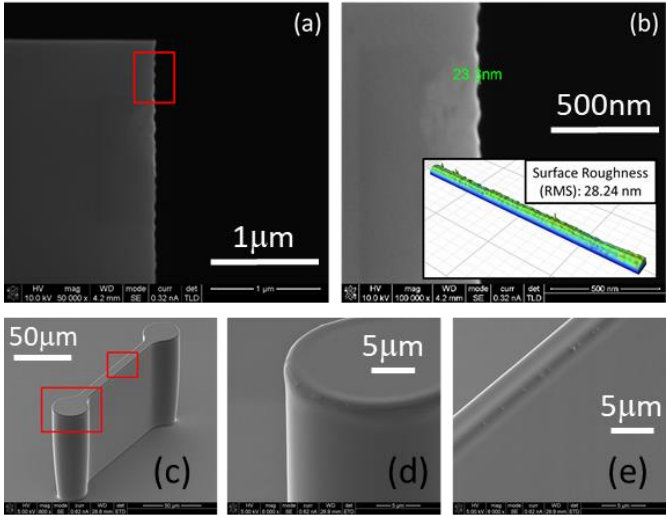

Fig. 1. The cross-sectional SEM of the fin highlighting (a) the rough sidewall after Si DRIE and (b) the sidewall profile and surface roughness measured by Bruker Optical Profilometer. (c-e) The SEM of Si micro-fins after $H_2$ annealing illustrating the smooth sidewall surfaces at different regions.

For the first and the second wafers (wafer 1 and wafer 2), the process is followed by the AC reactive magnetron sputtering of AlN directly on the Si micro-fins (i.e. without embedding a bottom metal layer). The third and the fourth wafers (wafer 3 and wafer 4) are used to study the effect of the bottom electrodes on the texture of sidewall AlN films. On the wafer 3, a 30 nm crystalline platinum (Pt) layer is atomically deposited in Cambridge Nano Fiji 200 Atomic Layer Deposition (ALD) tool at 150°C. On the wafer 4, sputtered molybdenum (Mo) is deposited, on a 20 nm AlN seed layer, to serve as the bottom electrode. Endeavor PVD cluster tool is employed [12] for the deposition of AlN and Mo thin films by AC and DC powered S-gun magnetrons, respectively. Two recipes are used for sputtering AlN on micro-fins in wafers 1-4. In both recipes, prior to AlN deposition, wafers are treated with RF plasma discharge at power of 70W ensuring effective argon (Ar) ion bombardment to atomically smoothen the fin surfaces [13][14]. This is followed by the AlN reactive sputtering at a base pressure of less than $2\times10^{-7}$ mbar with a power of 5.5kW. In the first recipe (recipe 1), the Ar and nitrogen ($N_2$) gas flows of 5 sccm and 17 sccm are used, respectively. In the second recipe (recipe 2), the Ar and $N_2$ gas flows are reduced to 3 sccm and 15 sccm, respectively. The recipe 1 is used for the wafer 1 and the recipe 2 is used for the wafer 2, and also wafers 3 and 4, after deposition of bottom electrode layer. The processing is continued on wafer 3, towards fabrication of operational FinBARs, by sputtering of 150 nm Mo layer with a DC power of 3kW. This is followed by patterning the top Mo electrode on the sidewall of the micro-fins, and finally creation of access windows to the bottom Pt electrode through etching of AlN using tetramethylammonium hydroxide (TMAH) solution at 50°C. The electrical admittances are calculated from the $S_{11}$ scattering parameters. To de-embed the effect of electrical feed-through on probing pads and parasitic resistance of the routing lines, planar calibration structures are used.

III. RESULTS AND DISCUSSION

A challenge with the characterization of sidewall AlN films is the limitations of the conventional X-ray diffraction (XRD) setup with large optical spot size that is optimized for the evaluation of planar films. To enable XRD analysis of the films on the minuscule sidewall area of the Si micro-fins, a sophisticated customization is required through coupling in-plane diffraction to synchrotron radiation with significantly reduced optical spot size. In the absence of the XRD results for sidewall AlN films, selected-area diffraction patterns, extracted from TEM images, are used [15]. A detailed set of bright-field cross-sectional transmission electron microscopy (BF-XTEM) images, taken across the sidewall film thickness, are used to identify the relative quality of the films over process variations, and also when compared with the films deposited on the planar surfaces in the same deposition run.

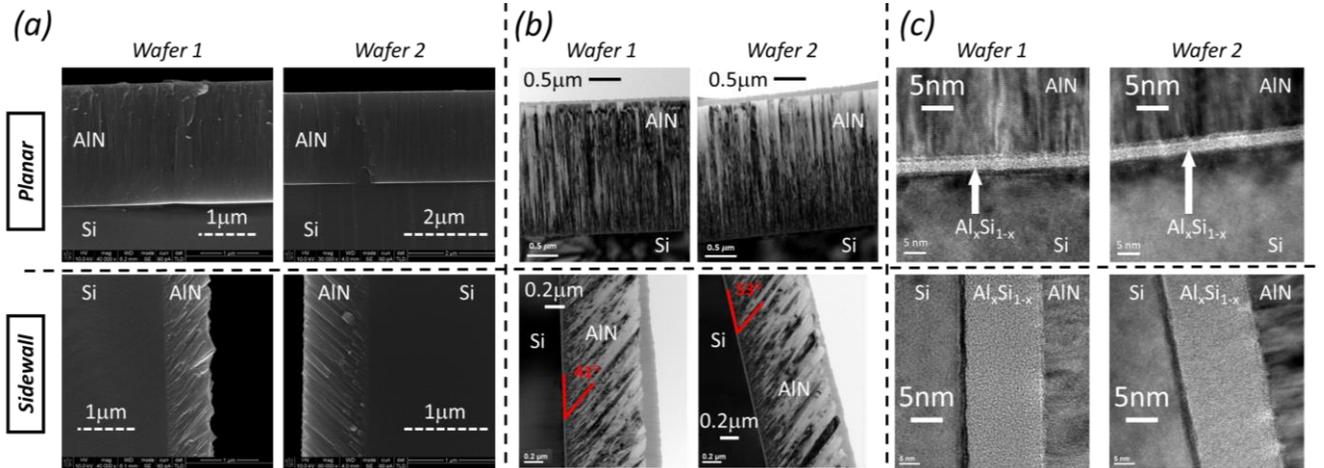

Figure 2. The (a) SEM and (b) TEM image of planar and sidewall AlN films deposited on wafers 1 and 2. (c) The TEM image at the interface with planar and sidewall Si surface highlighting the amorphous $Al_xSi_{1-x}$ layer.

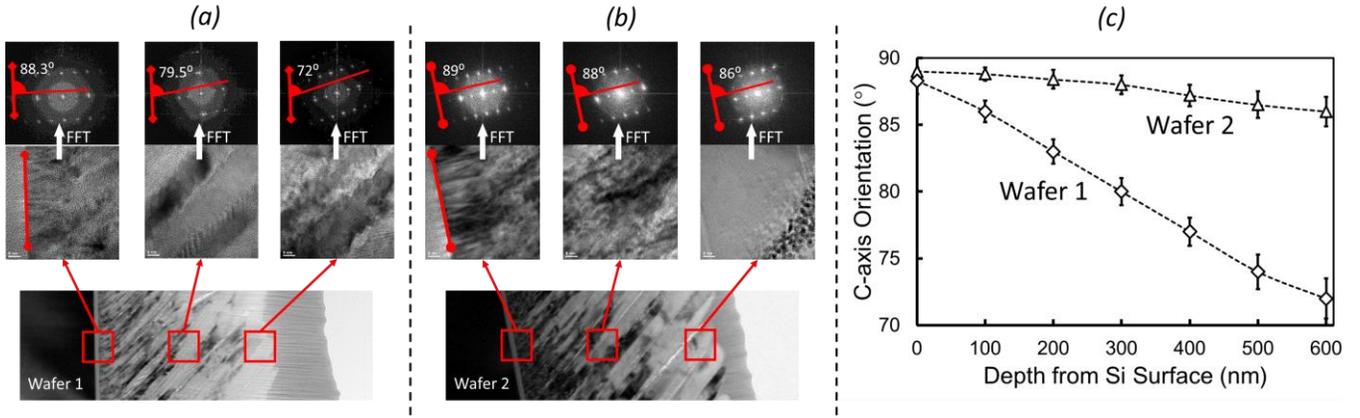

*Figure 3. The FFT of high-resolution XTEM extracted from TEM images for sidewall AlN films in (a) wafer 1 and (b) wafer 2. (c) Comparison of C-axis orientation over the sidewall AlN thickness for wafers 1 and 2. The data is collected over 20 different cross-lines.*

*A. Optimization of Sidewall AlN Deposition Process*

Figure 2 (a) compares the SEM images of the planar and sidewall AlN films for wafers 1 and 2. The two processes used for deposition of AlN on these wafers differentiated in the Ar and $N_2$ pressures. In both wafers, the thickness of sidewall films were nearly a third of the planar films. The slower deposition rate on the sidewall can be attributed to the geometric factor reducing flux of sputter species to the sidewall compared to a plane wafer surface [16][17]. Figure 2 (b) compares the TEM image of the planar and sidewall AlN films for wafers 1 and 2. From SEM and TEM images, it is evident that sputtering on sidewall results in tilted grains. The tilt in the sidewall AlN grains can be attributed to the increased directionality of the ad-atoms motion at lower gas flow to a preferred angle defined by the inherent architecture of the sputtering setup (i.e. the relative placement of wafer, target, and magnetron), and also the reduced mobility of the ad-atoms on the sidewall due to the increasing roughness of the film over the thickness [17-19]. These in turn correspond to the non-perpendicular direction of the deposition flux and slowed nucleation of the sidewall film that resulted in growth of thick amorphous aluminum silicide ($Al_xSi_{1-x}$) layer at the interface with Si surface. This can be clearly observed comparing the TEM images of the planar and sidewall films in both processes. While the $Al_xSi_{1-x}$ layer thickness is only 2 nm in planar films, its thickness increases to ~20 nm on the sidewall. The thicker amorphous $Al_xSi_{1-x}$ layer results in excessive roughness of the sidewall surface, which in turn promotes tilted growth in individual grains [20].

Comparing the SEM and TEM images for two processes it is evident that reduction in sputtering gas pressure significantly reduces the tilt angle of the grains. While the tilt angle of sidewall AlN grains in the process 1 is ~41°, it is ~53° for the process 2 with lowered deposition pressure. Figure 3 (a,b) compares the Fast Fourier transform (FFT) of the high-resolution XTEM, over selective locations across the thickness, for the sidewall films in wafers 1 and 2. The c-axis orientation, with respect to the surface is extracted for 20 locations uniformly distributed over the sidewall film thickness (Figure 3(c)). The c-axis orientation deviates from sidewall surface normal with thickness increase. The inconsistency and rotation of the sidewall AlN c-axis across the thickness results in the degradation of FinBAR $k_{eff}^2$ [6] and induces undesirable spurious modes in the frequency response of the resonator [17].

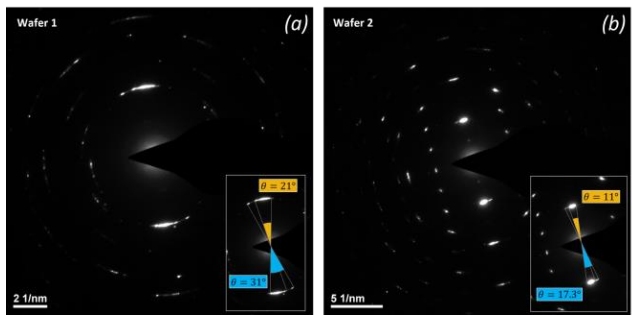

*Figure 4. The FFT of high-resolution diffraction patterns at the 300 nm depth of sidewall AlN films in (a) wafer 1 and (b) wafer 2. The inset images show the arc-angle around (002) spot.*

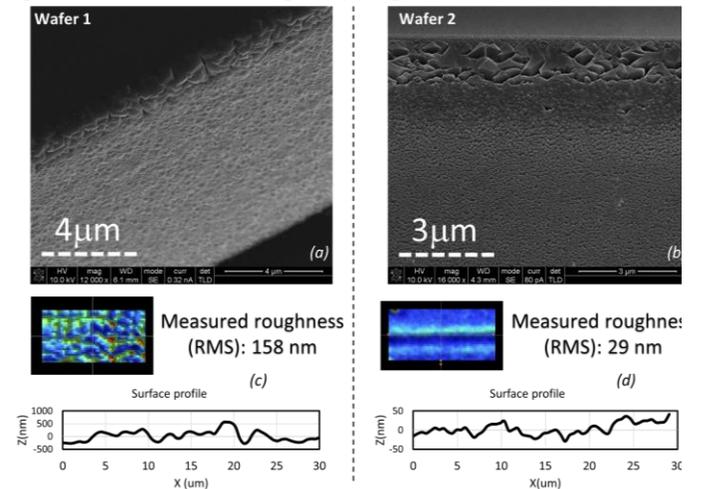

*Figure. 5. The SEM images of the sidewall AlN films deposited on (a) wafer 1 and (b) wafer 2. The measured surface roughness of the film for (c) wafer 1 and (d) wafer 2.*

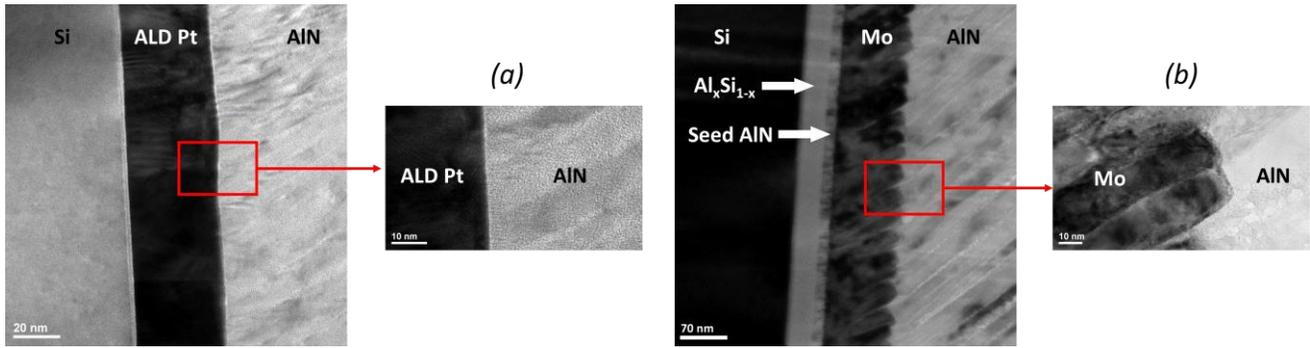

*Figure 6. The TEM images of the sidewall bottom electrode deposited on (a) wafer 3 and (b) wafer 4.*

C-axis orientations of 80.15°±8.15° for wafer 1 and 87.5°±1.5° for wafer 2 are extracted across the sidewall film thickness. This result highlights the significant improvement in normal orientation and cross-thickness consistency of sidewall AlN c-axis with the reduction of sputtering pressure. Furthermore, the lower sputtering pressure results in improved crystallinity of the sidewall film. Figure 4 compares the arc-angle around (002) spot extracted from the diffraction patterns of sidewall films at 300 nm depth, for wafers 1 and 2. Arc-angles of ~21°+10° and ~11°+6.3° are measured for the wafers 1 and 2, respectively, which highlights the improvement in sidewall film crystallinity with lowered sputtering pressure.

Finally, figure 5 shows the SEM and compares the surface roughness of the sidewall AlN films for wafers 1 and 2. The sidewall films have significantly higher roughness compared to planar counterparts in both wafers. This is due to the granular growth of sidewall films. Furthermore, the surface roughness is significantly decreased by reducing the sputtering pressure in process 2. While a surface roughness of 158 nm (rms) is measured on the sidewall film in wafer 1, reducing the sputtering pressure results in a surface roughness of 29 nm (rms) in wafer 2.

### B. Effect of Bottom Electrode

Following the optimization of the sputtering process on Si micro-fins, wafers 3 and 4 are used to explore the effect of different bottom electrodes on the texture and crystallinity of the sidewall films. Considering the higher quality of sidewall films sputtered at lower pressure, the process 2 is used for AlN deposition on wafers 3 and 4. It is well-known that addition of bottom electrode tremendously affects the quality of sputtered piezoelectric film [21-24]. The choice of the bottom electrode material and the deposition methodology is identified to ensure crystalline texture of the metallic film. In wafer 3, a 30 nm ALD Pt that is deposited on (110) Si shows a dominant (111) texture (0.1° FWHM on the top surface). In wafer 4, a seed AlN layer of ~20 nm is sputtered on the sidewall, using process 2, to promote (110)-crystalline growth of the bottom Mo layer. Figure 6 compares the TEM image of the sidewall bottom electrodes for wafers 3 and 4. While the ALD Pt has created a sharp interface with Si sidewall surface, the amorphous $Al_xSi_{1-x}$ layer with a thickness of 20 nm is evident in wafer 4 and results in granular growth of bottom Mo with large roughness. Figure 7 compares the TEM image of the subsequent AlN layer deposited on the bottom electrodes in wafers 3 and 4. A columnar structure is observed in wafer 3 with uniform grain sizes across the film thickness and over the sidewall area, which can be attributed to the smooth interface with Pt layer that enhanced the surface mobility of the adatoms [25]. In contrary, the excessive roughness of the bottom Mo layer in wafer 4 resulted in nonhomogeneous grain sizes. Consequently, while the wafer 3 with ALD Pt electrode shows a consistent crystalline texture across the sidewall film thickness, in wafer 4 with rough Mo electrode layer the quality of the sidewall AlN is substantially degraded. A c-axis orientation of 88.5°±1.5° and 78°±3° is measured for wafers 3 and 4, respectively. Besides, the arc-angle of 12°+2° and 24° are measured for wafers 3 and 4, respectively. While the quality of the sidewall film in wafer 4 is not suitable for implementation of FinBARs, the wafer 3 that utilize ALD Pt as the bottom electrode shows a comparable crystallinity to wafer 2 (no bottom electrode). Finally, figure 8 compares the surface roughness between wafers 2 and 3. While a similar c-axis orientation and crystallinity is observed in wafers 2 and 3, the addition of Pt bottom electrode has considerably reduced the surface roughness from 29 nm rms (wafer 2) to 16 nm rms (wafer 3). This improvement can be attributed to the effect of Pt layer as the diffusion barrier that prevents from formation of the rough and amorphous $Al_xSi_{1-x}$ layer at Si interface. Besides, the large surface roughness in wafer 4 samples can be attributed to the growth velocity anisotropy across the sidewall film thickness, which is induced by the inconsistency of the crystal orientation during the nucleation on the rough Mo layer [25].

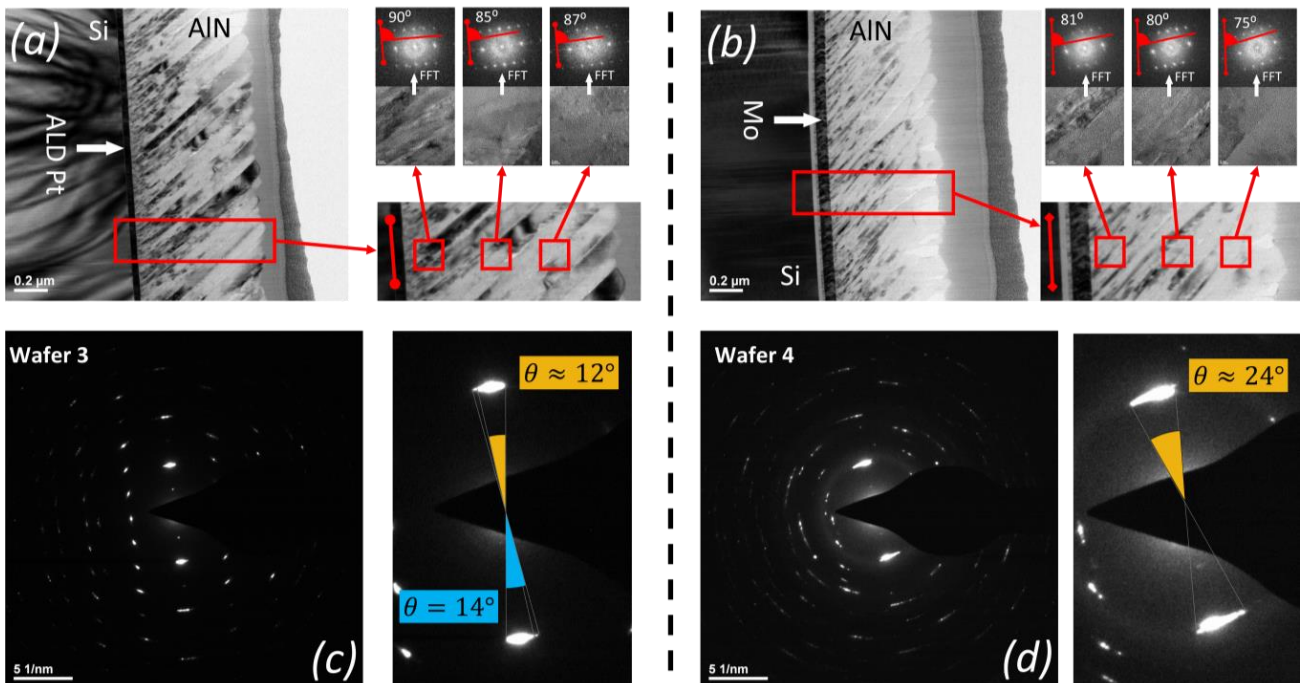

*Figure 7. The TEM image and FFT of diffraction patterns for the sidewall AlN film deposited on the bottom electrodes for (a) wafer 3 and (b) wafer 4.*

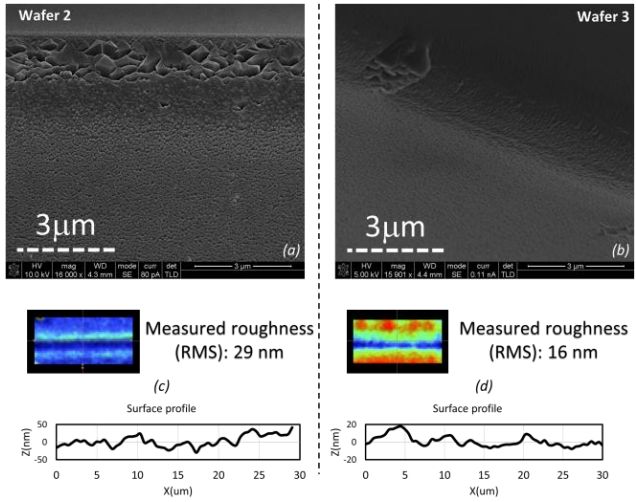

*Figure 8. The SEM images of the sidewall AlN films deposited on (a) wafer 2 and (b) wafer 3. The measured surface roughness of the film for (c) wafer 2 and (d) wafer 3.*

### C. FinBAR Characterization

FinBARs are fabricated on wafer 3 through deposition and patterning of sidewall Mo electrodes and opening access to bottom Pt electrode. Figure 9(a) shows the SEM image of a FinBAR with 2.2μm-wide fin, 30 nm ALD Pt as bottom electrode, 720 nm sidewall AlN, and 50 nm Mo as the top electrode on the sidewall.

Figure 9(b) shows the simulated admittance of the FinBAR around the 3rd width-extensional ($WE_3$) resonance mode at 4.35 GHz. The simulated cross-sectional mode shape of the $WE_3$ mode is shown as the inset, highlighting the efficient energy localization with the use of dispersion engineering of the lateral geometry [26][27]. The simulations are performed by assigning a low-reflectivity boundary condition at the bottom of the FinBAR to mimic the anchoring loss and its effect on the resonator $Q$. Figure 9(c) illustrates the measured admittance of the FinBAR, after de-embedding the static pad capacitance and routing resistances induced by unpatterned bottom Pt layer extended over the entire substrate. The FinBAR is operating in WE$_3$ mode at 4.23 GHz showing a $Q$ of 1,574 and $k_{eff}^2$ of 2.75%, which are both smaller compared to simulations that show a $Q$ of 2,600 (considering intrinsic acoustic dissipation in different materials and also the energy leakage into the substrate) and $k_{eff}^2$ of 3.78%.

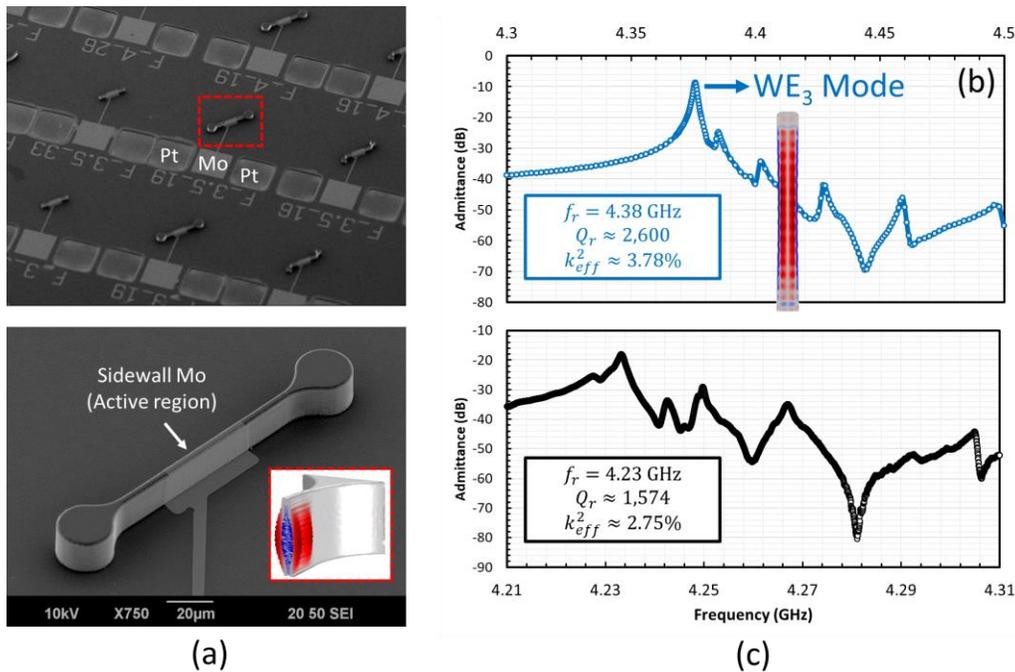

*Figure 9. (a) The SEM images of FinBARs implemented on the wafer 3. (b) The simulated and (c) the measured admittance of the FinBAR. The insets in (b) and (c) show the half-structure and cross-sectional simulated mode shapes of the FinBAR, respectively.*

The lower measured $k_{eff}^2$ and $Q$ compared to the simulated model of the FinBAR can be attributed to the lower quality of sidewall AlN film compared to the ideal case used in the simulation. Particularly, the structural nonideality of the developed sidewall film, which is composed of large tilted grains, induces undesirable scattering of the bulk acoustic waves and degrades both $k_{eff}^2$ and $Q$. Furthermore, the granular texture of the sidewall films results in excessive intragranular boundaries with nanocrystalline interface and further attenuates the acoustic wave as propagating within the structure.

## IV. CONCLUSION

This paper presents the process optimization of c-axis textured sidewall AlN films. These films enable realization of FinBAR technology that is poised to realize single-chip multi-band spectral processors. Highly c-axis oriented sidewall AlN films are achieved through low-pressure reactive magnetron sputtering. (111)-textured ALD Pt is used as the bottom electrode on the sidewall to suppress formation of amorphous Al$_x$Si$_{1-x}$ layer and facilitate crystalline growth of AlN with perpendicular c-axis orientation. This in turn reduced the surface roughness of the sidewall AlN film to ~16 nm. 4.2 GHz FinBARs fabricated based on the optimized sidewall AlN process shows a $Q$ of 1,574 and keff2 of 2.75% that are smaller compared to simulations. The lower $Q$ and $k_{eff}^2$ can be attributed to excessive intragranular surfaces in sidewall AlN that disperse the bulk acoustic vibration. Further work is needed to improve the quality of sidewall films to achieve FinBAR performances required for adoption in RF front-end applications. A proposed approach for improvement of the sidewall AlN quality is the application of substrate tilting and rotation during the film deposition, to suppress the geometrical factors responsible for the reduction of the sputter species flux on the sidewall surface.